\documentstyle[preprint,pra,aps]{revtex}
\begin{document}
\preprint{TAUP 2452-97}
\draft
\title{CAUSALITY AND NONLOCALITY AS AXIOMS FOR QUANTUM 
MECHANICS\thanks{For the proceedings of the symposium on {\it Causality and
Locality in Modern Physics and Astronomy:  Open Questions and Possible
Solutions}, York University, Toronto, August 25-29, 1997}}
\author{Sandu Popescu}
\address{University of Cambridge, Isaac Newton Institute, 20 Clarkson Road, 
Cambridge, U.K. CB3 0EH and BRIMS Hewlett-Packard Labs. Bristol BS12 6QZ,
U.K.}
\author{Daniel Rohrlich}
\address{School of Physics and Astronomy, Tel Aviv University, Ramat Aviv
69978 Israel}
\date{\today}
\maketitle
\begin{abstract}
Quantum mechanics permits nonlocality---both nonlocal correlations and
nonlocal equations of motion---while respecting relativistic causality.
Is quantum mechanics the unique theory that reconciles nonlocality and
causality?  We consider two models, going beyond quantum mechanics, of
nonlocality:  ``superquantum" correlations, and nonlocal ``jamming" of
correlations.  These models are consistent with some definitions of 
nonlocality and causality. 
\end{abstract}
\def\a{\hat a}
\def\ap{\hat a^\prime}
\def\b{\hat b}
\def\bp{\hat b^\prime}
\def\AP{A^\prime}
\def\BP{B^\prime}
\def \ra {\rangle}
\def \la {\langle}
\section{Introduction}
     ``But how can it be like that?"  This question, which every student
of quantum mechanics asks, is unanswerable, wrote Feynman; we should not
keep asking ourselves ``But how can it be like that?" lest we end up in a
blind alley from which no one has yet escaped.  According to Feynman, the
question reflects an utterly vain desire to see quantum mechanics in terms
of something familiar\cite{rpf}.  What exactly is the problem?  Is it that
an electron can be in two places at once?  No, we are used to such behavior
in electrons.  The problem is deeper.  We see the problem if we compare 
quantum mechanics with the special theory of relativity.  Special relativity
can be deduced in its entirety from two axioms:  the equivalence of inertial
reference frames, and the constancy of the speed of light.  Both axioms 
have clear physical meaning.  By contrast, the numerous axioms of quantum
mechanics have no clear physical meaning.  

     Shimony\cite{s1,s2} and Aharonov\cite{a1,a2} offer hope and a new
approach to this problem.  Their point of departure is the remarkable 
coexistence (peaceful or otherwise) of quantum nonlocality and the theory
of relativity.  Shimony focussed on the subtle nonlocality of quantum 
correlations.  On the one hand, quantum correlations themselves obey 
relativistic causality, in the sense that we cannot exploit quantum 
correlations to transmit signals at superluminal speeds\cite{c} (or at any 
speed).  On the one hand, as Bell\cite{bell} showed, quantum correlations 
could not arise in any theory of local hidden variables\cite{note1}.  
That quantum mechanics combines nonlocality and causality is wondrous.  
Nonlocality and causality seem {\it prima facie} incompatible.  Yet quantum 
correlations do not permit action at a distance, and Shimony\cite{s1} has 
aptly called the nonlocality manifest in quantum correlations ``passion at 
a distance".  Shimony has raised the question whether nonlocality and 
causality can peacefully coexist in any other theory besides quantum 
mechanics\cite{s1,s2}.

     Quantum mechanics also implies nonlocal equations of motion, as
Aharonov\cite{a1,a2} has pointed out.  In one version of the Aharonov-Bohm
effect\cite{ab}, a solenoid carrying an isolated magnetic flux, inserted
between two slits, shifts the interference pattern of electrons passing
through the slits.  The electrons therefore obey a nonlocal equation of
motion:  they never pass through the flux yet the flux affects their
final positions on the screen\cite{note2}.  Aharonov has shown that
the solenoid and the electrons exchange a physical quantity, the {\it
modular momentum}, nonlocally.  In general, modular momentum is measurable
and obeys a nonlocal equation of motion.  But when the flux is constrained
to lie between the slits, its modular momentum is completely uncertain,
and this uncertainty keeps us from seeing a violation of causality.  Nonlocal 
equations of motion imply action at a distance, but quantum mechanics just 
barely manages to respect relativistic causality.  Could it be, 
Aharonov\cite{a2} has asked, that quantum mechanics is the {\it unique} 
theory combining them?

     The parallel questions raised by Shimony and Aharonov lead us to
consider models for theories, going beyond quantum mechanics, that
reconcile nonlocality and causality.  Is quantum mechanics the only
such theory?  If so, nonlocality and relativistic causality together
imply quantum theory, just as the special theory of relativity can be
deduced in its entirety from two axioms\cite{a2}.  In this paper,
we will discuss model theories\cite{pr,gpr,rp} manifesting nonlocality
while respecting causality.  The first model manifests nonlocality in
the sense of Shimony:  nonlocal correlations.  The second model
manifests nonlocality in the sense of Aharonov:  nonlocal dynamics.
These models raise new theoretical and experimental possibilities.  
Apparently, quantum mechanics is {\it not} the only theory that 
reconciles nonlocality and relativistic causality.  Yet, it is 
possible that stronger axioms of locality and causality could rule 
out both models.  Thus we have no final answer to the question, Is 
quantum mechanics the only theory that reconciles nonlocality and
causality?  It is not surprising that we cannot offer a final answer to the 
question; it is perhaps surprising that we can offer any answer.  But most 
of all, we offer the {\it attempt} to formulate and answer the question.

\section{Nonlocality I:  nonlocal correlations}
The Clauser, Horne, Shimony, and Holt\cite{chsh} form of Bell's
inequality holds in any classical theory (that is, any theory of local
hidden variables).  It  states that a certain combination of correlations
lies between -2 and 2:
\begin{equation}
-2 \le E(A,B)+E(A,\BP )+E(\AP ,B)-E(\AP ,\BP )\le 2
~~~~.
\label{1}
\end{equation}
Besides 2, two other numbers, $2\sqrt{2}$ and $4$, are important bounds
on the CHSH sum of correlations.  If the four correlations in Eq.\ (\ref{1})
were independent, the absolute value of the sum could be as much as 4. For
quantum correlations, however, the CHSH sum of correlations is bounded
\cite{t} in absolute value by $2\sqrt{2}$. Where does this bound come from?
Rather than asking why quantum correlations violate the CHSH inequality,
we might ask why they do not violate it {\it more}.  Suppose that quantum
nonlocality implies that quantum correlations violate the CHSH inequality
at least sometimes.  We might then guess that relativistic causality is
the reason that quantum correlations do not violate it maximally. Could
relativistic causality restrict the violation to $2\sqrt {2}$ instead of
4?  If so, then nonlocality and causality would together determine the
quantum violation of the CHSH inequality, and we would be closer to a
proof that they determine all of quantum mechanics.  If not, then quantum
mechanics cannot be the unique theory combining nonlocality and causality.
To answer the question, we ask what
restrictions relativistic causality imposes on joint probabilities.
Relativistic causality forbids sending messages faster than light. Thus,
if one observer measures the observable $A$, the probabilities for the
outcomes $A=1$ and $A=-1$ must be independent of whether the other observer
chooses to measure $B$ or $\BP$. However, it can be shown\cite{pr,etc}
that this constraint does not limit the CHSH sum of quantum
correlations to $2\sqrt{2}$. For example, imagine a ``superquantum"
correlation function $E$ for spin measurements along given axes.
Assume $E$ depends only on the relative angle $\theta$ between axes.
For any pair of axes, the outcomes $\vert \uparrow \uparrow \rangle$
and $\vert \downarrow \downarrow \rangle$ are equally likely, and
similarly for $\vert \uparrow \downarrow \rangle$ and $\vert \downarrow
\uparrow \rangle$.  These four probabilities sum to 1, so the probabilities
for $\vert \uparrow \downarrow \rangle$ and $\vert \downarrow \downarrow
\rangle$ sum to $1/2$. In any direction, the probability of $\vert
\uparrow \rangle$ or $\vert \downarrow \rangle$ is $1/2$ irrespective
of a measurement on the other particle.  Measurements on one particle
yield no information about measurements on the other, so relativistic
causality holds.  The correlation function then satisfies $E(\pi -
\theta ) = -E(\theta )$.  Now let $E(\theta )$ have the form

     (i) $E (\theta ) =1$ for $0 \le \theta \le \pi /4$;

     (ii) $E(\theta )$ decreases monotonically and smoothly from 1 to -1
as $\theta$ increases from $\pi /4$ to $3\pi / 4$;

     (iii) $E(\theta) = -1$ for $3\pi /4 \le  \theta \le \pi$.

Consider four measurements along axes defined by unit vectors $\ap$, $\b$,
$\a$, and $\bp$ separated by successive angles of $\pi /4$ and lying in a
plane. If we now apply the CHSH inequality Eq.\ (\ref{1}) to these
directions, we find that the sum of correlations
\begin{equation}
E(\a , \b ) +E(\ap , \b )+E(\a , \bp )-E(\ap , \bp )
=3E(\pi /4 ) - E(3\pi /4) = 4
\end{equation}
violates the CHSH inequality with the maximal value 4.  Thus, a correlation
function could satisfy relativistic causality and still violate the CHSH
inequality with the maximal value 4.

\section{Nonlocality II: nonlocal equations of motion}
     Although quantum mechanics is not the unique theory combining
causality and nonlocal correlations, could it be the unique theory
combining causality and nonlocal equations of motion?  Perhaps the
nonlocality in quantum dynamics has deeper physical signficance.  Here
we consider a model that in a sense combines the two forms of
nonlocality:  nonlocal equations of motion where one of the physical
variables is a nonlocal correlation.  {\it
Jamming}, discussed by Grunhaus, Popescu and Rohrlich\cite{gpr} is such
a model.  The jamming paradigm involves three experimenters.  Two
experimenters, call them Alice and Bob, make measurements on systems that
have locally interacted in the past.  Alice's measurements are spacelike
separated from Bob's.  A third experimenter, Jim (the jammer), presses a
button on a black box.  This event is spacelike separated from Alice's
measurements and from Bob's.  The black box acts at a distance on the
correlations between the two sets of systems.  For the sake of definiteness,
let us assume that the systems are pairs of spin-1/2 particles entangled
in a singlet state, and that the measurements of Alice and Bob yield
violations of the CHSH inequality, in the absence of jamming; but when
there is jamming, their measurements yield classical correlations (no
violations of the CHSH inequality).  Indeed, Shimony\cite{s1} considered
such a paradigm in the context of the experiment of Aspect, Dalibard, and
Roger\cite{adr}; his concern was to probe hidden-variable theories due
to Vigier and others\cite{hv}.  

     Here, we ask whether such a nonlocal equation
of motion (or one, say, allowing the third experimenter nonlocally to
create, rather than jam, nonlocal correlations) could respect causality.
The jamming model\cite{gpr} addresses this question. In general,
jamming would allow Jim to send superluminal signals.  But remarkably,
some forms of jamming would not; Jim could tamper with nonlocal
correlations without violating causality.  Jamming preserves causality
if it satisfies two constraints, the {\it unary} condition and the {\it
binary} condition.  The unary condition states that Jim cannot use
jamming to send a superluminal signal that Alice (or Bob), by examining
her (or his) results alone, could read.  To satisfy this condition,
let us assume that Alice and Bob each measure zero average spin along
any axis, with or without jamming. In order to preserve causality,
jamming must affect correlations only,
not average measured values for one spin component.  The binary condition
states that Jim cannot use jamming to send a signal that Alice {and}
Bob {\it together} could read by comparing their results, if they could
do so in less time than would be required for a light signal to reach
the place where they meet and compare results.  This condition restricts
spacetime configurations for jamming.  Let $a$, $b$ and $j$ denote the
three events generated by Alice, Bob, and Jim, respectively: $a$ denotes
Alice's measurements, $b$ denotes Bob's, and $j$ denotes Jim's pressing of
the button.  To satisfy the binary condition, the overlap of the forward
light cones of $a$ and $b$ must lie entirely {\it within} the forward light
cone of $j$.  The reason is that Alice and Bob can compare their results
only in the overlap of their forward light cones.  If this overlap is
entirely contained in the forward light cone of $j$, then a light signal
from $j$ can reach any point in spacetime where Alice and Bob can compare
their results.  This restriction on jamming configurations also rules
out another violation of the unary condition.  If Jim could obtain the
results of Alice's measurements prior to deciding whether to press the
button, he could send a superluminal signal to Bob by {\it selectively}
jamming\cite{gpr}.

      An odd feature in this model is that, in principle, it allows the
effect (the correlations measured at $a$ and $b$) to precede the effect
(the action of Jim at $j$).  Such reversals may boggle the mind, but 
they do not lead to any inconsistency as long as they do not generate 
self-contradictory causal loops\cite{dbohm,a3}.  It is not hard to 
show\cite{gpr} that if jamming satisfies the unary and binary conditions,
it does not lead to self-contradictory causal loops, regardless of the 
number of jammers.  While we argue that jamming is consistent even if 
it allows reversals of the sequence of cause and effect, we also point
out that such reversals arise only in one space dimension.  In higher 
dimensions, jamming is not possible if both $a$ and $b$ precede $j$;
the binary condition itself eliminates such configurations\cite{shim}. 

\section{Conclusions}
     Two related questions of Shimony\cite{s1,s2} and Aharonov\cite{a2}
inspire our work.  Nonlocality and relativistic causality seem {\it
almost} irreconcilable.  The emphasis is on {\it almost}, because
quantum mechanics does reconcile them, and does so in two different
ways.  But is quantum mechanics the unique theory that does so?  Our
preliminary answer is that it is not:  model theories going beyond 
quantum mechanics, but respecting causality, allow nonlocality.  
However, we qualify our answer.  First, nonlocality is not completely 
defined.  Nonlocality in quantum mechanics includes both nonlocal 
correlations and nonlocal equations of motion, and we do not know exactly
what kind of nonlocality we are seeking.  Second, even causality is not
completely defined; there are formulations of causality that we often
take to be equivalent, that are not strictly equivalent.  
Is quantum mechanics the unique theory that reconciles nonlocality
and causality?  We cannot offer a final answer to the question, but we 
offer the {\it attempt} to formulate and answer the question.  Whether 
by strengthening the axioms of nonlocality and causality, or by adding 
new axioms with clear physical meaning, we hope to rediscover quantum 
mechanics as the unique theory consistent with these axioms.

\acknowledgments
D. R. acknowledges support from the State of Israel, Ministry of Immigrant
Absorption, Center for Absorption in Science (Giladi Program).


\begin{references}
\bibitem{note2}
It is true that the electron interacts {\it locally} with
a vector potential.  However, the vector potential is not a physical
quantity; all physical quantities are gauge invariant.
\end{references}
\end{document}